\documentclass[12pt,a4paper]{article}
\usepackage{a4wide}
\usepackage{latexsym}
\usepackage{epsf}
\usepackage{amssymb}
\begin{document}
\def\be{\begin{equation}}
\def\ee{\end{equation}}
\def\bea{\begin{eqnarray}}
\def\eea{\end{eqnarray}}
\def\ba{\begin{array}}
\def\ea{\end{array}}

\def\pd{\partial}
\def\a{\alpha}
\def\b{\beta}
\def\g{\gamma}
\def\d{\delta}
\def\m{\mu}
\def\n{\nu}
\def\t{\tau}
\def\l{\lambda}
\def\th{\theta}
\def\O{\Omega}
\def\r{\rho}
\def\s{\sigma}
\def\e{\epsilon}
\def\x{\xi}
\def\T{\Theta}
\def\O{\Omega}
\def\scri{\mathcal{J}}
\def\cM{\mathcal{M}}
\def\tcM{\tilde{\mathcal{M}}}
\def\RR{\mathbb{R}}

\hyphenation{re-pa-ra-me-tri-za-tion}
\hyphenation{trans-for-ma-tions}


\begin{flushright}
IFT-UAM/CSIC-03-10\\
hep-th/0303164\\
\end{flushright}

\vspace{1cm}

\begin{center}

{\bf\Large The Dirichlet Obstruction in AdS/CFT }

\vspace{.5cm}

{\bf Enrique \'Alvarez, Jorge Conde and Lorenzo Hern\'andez }

\vspace{.3cm}

\vskip 0.4cm

{\it  Instituto de F\'{\i}sica Te\'orica UAM/CSIC, C-XVI,
and  Departamento de F\'{\i}sica Te\'orica, C-XI,\\
  Universidad Aut\'onoma de Madrid 
  E-28049-Madrid, Spain }

\vskip 0.2cm

\vskip 1cm

{\bf Abstract}
The  obstruction for
a perturbative  reconstruction of the five-dimensional bulk metric 
starting from the four-dimensional metric at the boundary,
that is, the Dirichlet 
problem, is computed in dimensions $6\leq d\leq 10$ and some comments are made 
on its general structure and, in particular, on its relationship 
with the conformal anomaly, which we compute in dimension $d=8$.

\end{center}

\begin{quote}

\end{quote}


\newpage

\setcounter{page}{1}
\setcounter{footnote}{1}
\newpage
\section{Introduction}
One of the more interesting problems in the celebrated AdS/CFT correspondence
conjectured by Maldacena (\cite{malda}) in the language introduced by
 Witten (\cite{witten}; cf. also  the reviews \cite{aharony},\cite{Alvarez}, \cite{Petersen})
 is the {\em decoding of the hologram }. In the minimal
gravitational
setting this amounts to recover the metric in the bulk space from the
metric at the boundary (which is what mathematicians call a {\em Dirichlet 
problem}).
\par
To be specific, in the framework of the geometric approach to holography  
in its Poincar\'e
 form (that is, when the d-dimensional manifold $M_d$ is
represented as Penrose's conformal infinity of another manifold $B_{d+1}$
with one extra holographic dimension),
there is a privileged system of coordinates such that the {\em boundary}
$\pd B_{d+1}\sim M_d$ is located at the value $\rho = 0$ of the holographic coordinate, namely
\be
ds^2 = {g^{(B)}}_{\m \n}dx^{\m}d^{\n} = \frac{-l^2 d\rho^2}{4\rho^2} + \frac{1}{\rho} h_{ij}(x,\rho)dx^i dx^j
\ee\label{hs}
(The normalization corresponds to a cosmological constant 
$\lambda\equiv \frac{d-2}{2d} R\equiv \frac{d(d-1)}{2 l^2}$ when $h_{ij}
=\eta_{ij}$).
It has been already mentioned that the boundary condition is of the Dirichlet type, i.e.
\be
h_{ij}(x,\rho = 0) = g_{ij}(x)
\ee
where $g_{ij}$ is an appropiate metric on $M_d$.
\par
\par
 In the
 basic mathematical work by Fefferman and Graham (FG) (\cite{fefferman}) ,
it is proved that 
there is a formal power
series solution to Einstein's equations with a cosmological constant as above.
This power series gives, in principle, a complete solution to the 
Dirichlet problem, at least in the vicinity of the boundary.
When $d\in 2\mathbb{Z}$ there is, however, an obstruction, which in 
four dimensions is the Bach tensor
 and in higher dimensions 
is a new tensor, whose specific form is unknown, which will be called in 
the sequel the FG tensor, $Z^d_{ij}$, with conformal weight 
$(d-2)/2$. When this tensor does not vanish, 
there are logarithmic terms which appear in the expansion starting at the
order $\r^{d/2}$ 
\be\label{metric}
h_{ij} = g_{ij} +h^{(1)}_{ij} \rho  + \ldots+ h^{(d/2)}_{ij}\rho^{d/2}  + 
\tilde{h}^{(d/2)}_{ij} \rho^{d/2} \log{\rho} +\ldots
\ee
Even the term $h^{(d/2)}_{ij}$ is not 
completely determined; as shown in \cite{haro}, Einstein's
equations only give its trace as well as its covariant
derivative. On the other hand, the coefficient of the logarithmic
term is given in terms of the local anomaly  $a_d$ by (cf. \cite{haro}).
\be\label{haro}
\tilde{h}_{ij}^{d/2}=-\frac{4}{d\sqrt{g}}\frac{\delta}{\delta g^{ij}}\int dx_d
a_{d}
\ee 
i.e., it is the energy momentum tensor of the integrated anomaly.

\par
The object of the present paper is to analyze this problem in detail, using
in particular some codimension two techniques, also due to FG originally. We shall find general 
expressions for the obstructions, recursive relationships between them, and
explicit formulas for the eight dimensional conformal anomaly.
\section{Einstein's equations for $B_{d+1}$ and $\lambda\neq 0$}

 Einstein's equations \footnote{
We use the conventions $R^{\m}_{\n \a \b} = 2\pd_{[ \a} \Gamma^\m _{ \b ] \n} + 2\Gamma^\m _{\s [\a} \Gamma^\s _{ \b ] \n}$ and $R_{\m \n} = R^\a _{\m \a \n}$.}
corresponding to a cosmological constant  $\lambda = \frac{d(d-1)}{2l^2}$:
read
\be
R_{\m \n} = \frac{d}{l^2} {g^{(B)}}_{\m \n}
\ee
Which reduce in the FG coordinates to:
\bea \label{EOMS}
\label{EOM1}
\rho \left( 2 h'' - 2h'h^{-1}h' + tr[h^{-1}h']h'\right)_{ij} + l^2 R[h]_{ij} - (d-2)h'_{ij} - tr[h^{-1}h']h_{ij} =0\\
\label{EOM2}
(h^{-1})^{kl}(\nabla^{[h]}_i h'_{kl} - \nabla^{[h]}_l h'_{ik}  ) =0 \\
\label{EOM3}
tr[h^{-1}h''] - \frac{1}{2}tr[h^{-1}h'h^{-1}h'] =0 
\eea
We shall consistently denote by $g_{ij}\equiv h_{ij}(\r=0)$  the (Penrose) boundary 
metric and from now on latin indices will be raised and lowered using this
 metric\footnote{We shall denote by $h^{(n)}\equiv
  g^{ij}h^{(n)}_{ij}$.}. Equations (\ref{EOM1}) and (\ref{EOM3}) (the ones
needed for our purposes) can now be expanded in a formal power series of
which we give the first two orders.\\
Order $\r^0$.
\bea \label{0order}
&& \bullet\quad (d-2) h^{(1)}\,_{ij} +h^{(1)} g_{ij} -l^2 {R[g]}_{ij} =0
   \label{0order1}\\
&&\bullet\quad 2h^{(2)} -\frac{1}{2} {h^{(1)}}^{kl} {h^{(1)}}_{kl} =0
   \label{0order2}
\eea
Order $\r^1$. 
\bea \label{1order}
&&\bullet\quad 2(d-4){h^{(2)}}_{ij} + \left( 2 h^{(2)} 
        -{h^{(1)}}^{kl}{h^{(1)}}_{kl}\right) g_{ij} +
        2{{h^{(1)}}_{i}}^{m}{h^{(1)}}_{jm}\nonumber\\
&& \quad- \frac{l^2}{2}\left[\nabla_{k}\nabla_i {{h^{(1)}}_j}^k 
    + \nabla_{k}\nabla_j {{h^{(1)}}_i}^k 
   - \nabla^2 {h^{(1)}}_{ij}-\nabla_{i}\nabla_{j}{h^{(1)}}\right] =0
   \label{1order1}\\
&&\bullet\quad 6h^{(3)} +
{{h^{(1)}}^{k}}_{l}{{h^{(1)}}^{l}}_{n}{{h^{(1)}}^{n}}_{k} - 4
{h^{(1)}}^{kl}{h^{(2)}}_{kl} =0 \label{1order2}
\eea
Explicit expressions up to order $\rho^4$ can be found in the Appendix \ref{HighEOMS}.
\section{ The Dirichlet obstruction in d=4 and d=6}
%
%
Equations (\ref{0order1}) and (\ref{0order2}) give $h^{(1)}_{ij}$ and $h^{(2)}$
\bea
\label{h1}
&&h^{(1)}_{ij} = \frac{l^2}{d-2} \left(R_{ij} - \frac{R}{2(d-1)}g_{ij}\right)
 \equiv l^2 A_{ij}\\
&&h^{(1)} = l^2 \frac {R}{2(d-1)} = l^2 A\\
&&h^{(2)} = \frac{1}{4} h^{(1)kl}h^{(1)}_{kl} = \frac{l^4}{4(d-2)^2} 
\left( R^{kl}R_{kl} + \frac{4-3d}{4(d-1)^2} R^2 \right)= \frac{l^4}{4} A^{kl}A_{kl}
\eea

If we now remember the definition of the  Bach tensor (using the conventions in \cite{Alvarez})
\be
B_{ij} \equiv 2\nabla^{k}\nabla_{[k}A_{i]j} + A^{kl} W_{kijl} 
\ee
(where $W_{ijkl}$ is the Weyl tensor), and use it in equation (\ref{1order1}), we get
%
%
%
%
\be
(d-4)(4h^{(2)}_{ij} - h^{(1)}_{im} h^{(1)}_{j}\, ^{m}) + l^4 B_{ij} =0
\ee
conveying the fact that, in $d=4$, if $B_{ij} \neq 0$ the expansion is 
inconsistent. ?????Note that, had we introduced the logarithmic term, a new coefficient in the 
expansion of the metric comes into play and we get a chance to make  Einstein's
 equations vanish up to second order, as FG claim. 

On the other hand, if the Bach tensor vanishes, the only further information 
of $h^{(2)}$ given up to this order is the value of its covariant derivative, 
which can be obtained from equation  (\ref{EOM2}).
\par

It is interesting to point out  that there is a very simple 
relationship
 between the Bach tensor and the four-dimensional anomaly.
\par
The vanishing of the Bach tensor \cite{bach} is the equation of motion 
corresponding
to the lagrangian:
\bea\label{weyl}
&&L\equiv I_4\equiv \frac{1}{64}\sqrt{g}W_{\alpha\beta\gamma\delta}W^{\alpha\beta\gamma\delta}\nonumber\\
&&= \frac{1}{64}\sqrt{g}(R^{\alpha\beta\gamma\delta}R_{\alpha\beta\gamma\delta} - 
2 R^{\alpha\beta}R_{\alpha\beta} + \frac{1}{3} R^2)
\eea
But precisely the anomaly (modulo local counterterms) for conformal invariant matter (\cite{Deser}) is given by a 
combination of the Euler characteristic density, $E_4$ 
plus this conformal invariant lagrangian, $I_4$. In four dimensions the Euler
characteristic density, is given explicitly by
\be
E_4=\frac{1}{64}\epsilon^{abcd}R_{ab\mu\nu}R_{cd\rho\sigma}\epsilon^{\mu\nu\rho\sigma}
\equiv \frac{1}{64} (R^{\alpha\beta\gamma\delta}R_{\alpha\beta\gamma\delta} - 
4 R^{\alpha\beta}R_{\alpha\beta} + R^2)
\ee
and is a topological invariant, that is, independent of the metric.
The vanishing of the Bach tensor is then the condition for the integrated 
anomaly to be topological, that is, independent of the metric,
 which according
to the previous equation (\ref{haro}) is the condition for 
\be
\tilde{h}_{ij}^{d/2}=0
\ee
in which case there is no obstruction for a perturbative solution to the
 Dirichlet problem.
\par
The fact that
\be
Z^{(d)}_{ij}\sim\tilde{h}_{ij}^{d/2}
\ee
gives (using (\ref{haro})) the FG tensor as the variational derivative of the conformal 
anomaly with respect to the background metric. 
Although both quantities are unknown in the general case, this is a 
 useful constraint.
\par
As an example, we shall begin with the six dimensional case. As in $d=4$, 
the obstruction comes out taking the limit $d \longrightarrow 6$ in the 
left hand side of the first equation in (\ref{2order}), that is:
\bea\label{obstruction}
H_{ij}&=& \frac{1}{4}(\frac{1}{2}B_{kl}A^{kl}+A^k_l A _m A ^m _k)g_{ij} -B^m _{(i}A_{j)m} + \frac{1}{8}AB_{ij} - \frac{1}{4}AA_{im}A_j ^m  \nonumber \\
&&+\frac{1}{8}\nabla_k \left(-\frac{1}{2} {DB_{i j}}^k \ _j + 2A_{l(i}C_{j)}\
  ^{kl} - 3A^{k}_l {DA_{i j}}^l + A^{kl}\nabla_{l}A_{ij} \right)  \nonumber \\
&&+\frac{1}{4} A^{kl}\nabla_i \nabla_j A_{kl} - \frac{1}{4}\nabla_l {DA_{i j}}^l + \frac{1}{2}\nabla_l A_{ik} C^{l k}\ _j 
\eea
where $ C_{i j k}=2\nabla_{[i}A_{j] k} $ is the Cotton tensor.
Then, according to our previous reasoning,
this obstruction should coincide with the
metric variation of the anomaly.
\par
To apply a first check to this idea, let us recall the expression for the six-dimensional
 holographic anomaly that has been proposed in (\cite{Henningson}):
\be
\mathcal{A} \sim \frac{1}{2}RR^{ij}R_{ij} - \frac{3}{50}R^3 - R^{ij}R^{kl}R_{ikjl} + \frac{1}{5}R^{ij} \nabla_i \nabla_j R -\frac{1}{2}R^{ij} \Box R_{ij} + \frac{1}{20}R \Box R
\ee
which can also be written as:
\be
\mathcal{A} \sim A^{ij} ( B_{ij} + 2A_{ik}A_j ^k - 3AA_{ij} + A^2 g_{ij})
\ee

We are interested in the metric variation, that is,
\be
\tilde{H}_{ij} \equiv \frac{\delta}{\delta g^{ij}}\int \sqrt{g} \mathcal{A} 
\ee
yielding:
\bea
\tilde{H}_{ij}= -\frac{1}{2}B^{kl}A_{kl}g_{ij}+\frac{1}{2}AB_{ij}+3A_{k(i}B_{j)}^k -A_{(i}^s A^{kl}W_{j)kls} +2(A^{kl}A-A^{ks}A^l_s)W_{kijl}  \nonumber\\
+A^{kl} A_{kl}A_{ij} +5AA_{ik}A_{j}^k - 6 A_{kl}A^k_i A^l_j -(A_{kl}A^{kl}A -
A^k_sA^{ls}A_{kl})g_{ij}+\frac{1}{4}\nabla_k {DB_{i j}}^k \nonumber\\
+\nabla^2(AA_{ij}-A_i^sA_{js}) - \nabla_k \nabla_l (A^{kl}A_{ij}-A^k_{(i} A^l_{j)}) -\frac{1}{2}\nabla_i \nabla_j(A^{kl}A_{kl} +A^2) \nonumber\\
-\nabla^k( A_{(i}^lC_{j)lk} + A_{(i}^lC_{j)kl} + A^{kl}C_{l(ij)}) +\nabla_{(i} A^{kl}C_{j)kl} +2\nabla_{(i|}(A^{kl}\nabla_k A_{|j)l})
\eea

As a matter of fact, the obstruction given by (\ref{obstruction}), can be manipulated 
in order to prove, after some calculation, that both tensors $ - 4 H$ and
$\tilde{H}$ are identical.

\section{ A general codimension-two approach}\label{C2}
In order to work out higher dimensions, it proves useful to put at work the codimension two approach
of (\cite{fefferman} , reviewed in  \cite{Alvarez}).
In this approach, the conformal invariants of the manifold $M_n$ are obtained from
diffeomorphism invariants of the so-called {\em ambient space}, $A_{n+2}$ with two extra dimensions,
one spacelike and the other timelike (cf. the analysis of
diffeomorphisms that reduce to Weyl transformations on the boundary (PBH) in \cite{Alvarezz}).
\par
If the ambient space is to have lorentzian signature, then the boundary is neccessarily euclidean.
Then, of the two extra ({\em holographic}) coordinates, one is timelike, and the other, spacelike.
They will be denoted by $(t,\r)$, respectively.
\par
In the references just cited it is proven that one can choose coordinates in such a way
that the ambient metric reads 
\be
ds^{2}={g^{(A)}}_{\m \n}dx^{\m}dx^{\n} = \frac{t^2}{l^2}h_{ij}(x,\r)dx^{i}dx^{j}+\r dt^{2}+t dt d\r
\ee
A power expansion in the variable $\r$ can now be performed:
\be
h_{ij}(x,\r)=g_{ij}(x)+\sum_{a=1}^{\infty}\r^a {h^{(a)}}_{ij}(x)
\ee
($g_{ij}$ is the boundary metric tensor, and it is assumed that the dimension of this space is even). 

The ambient space has to be Ricci-flat  (this is a necessary 
ingredient of the FG construction, which also guarantees that it is an
admissible string background). This means that
\be
\stackrel{^{(A)}}{\mathcal R}\,_{ij}=\rho \left( 2h'' -2h'h^{-1}h' +tr[h^{-1}h']h'\right)_{ij} +l^2 R[h]_{ij} -(d-2)h'_{ij} -tr[h^{-1}h']h_{ij} =0 
\ee
which is identical to the equation (\ref{EOM1}), as has been proved in detail 
in \cite{Alvarezz}.
It is remarkable that the Ricci flatness condition in $A_{n+2}$ is indeed
equivalent to Einstein's equations with an appropiate cosmological constant in the associated bulk space, $B_{n+1}$. 

Those equations are written for arbitrary  dimension, but it is clear just by
looking at the equivalent form (\ref{0order}), etc., that for  any even dimension $(d=2n)$  the equations are inconsistent unless the tensor that is not multipled by $(d-2n)$ vanishes, so that it is natural to call that tensor {\em the obstruction}.
  Let us insist that even when the obstruction vanishes
 there is an indetermination in the expansion from 
$h^{(n)}$ on.
\par

We turn our attention now to tensors defined in the ambient space. For
instance, the non vanishing components of the Riemann tensor are
\be 
\begin{array}{l}
{\mathcal R}_{ijkn}=\frac{t^2}{l^2}\left( 
R_{ijkn}(h)+\frac{1}{l^2}[-h_{ik}\pd_{\r}h_{jn}+h_{in}\pd_{\r}h_{jk}-h_{jn}\pd_{\r}h_{ki}+h_{jk}\pd_{\r}h_{in}]\right.
\nonumber \\
\hspace{1.2cm}\left. +\r \frac{1}{l^2}[\pd_{\r}h_{ik}\pd_{\r}h_{jn}-\pd_{\r}h_{in}\pd_{\r}h_{jk}]\right) \nonumber \\
{\mathcal R}_{\r ijk}=\frac{t^2}{l^2}\frac{1}{2}(\stackrel{
  ^{(h)}}{\nabla_{k}}\pd_{\r}h_{ij} -\stackrel{ ^{(h)}}{\nabla_{j}}\pd_{\r}h_{ik})\nonumber \\
{\mathcal R}_{\r i \r j}=\frac{t^2}{l^2}\frac{1}{2}(\pd_{\r}\pd_{\r}h_{ij}-\frac{1}{2}h^{kl}\pd_{\r}h_{kj}\pd_{\r}h_{li})
\end{array}
\ee

If we take the $\r \to 0$ limit we obtain
\be \label{riemann0} \ba{l}
R_{\r i \r j}=-\frac{t^{2}}{2l^{2}}(2{h^{(2)}}_{ij}-\frac{1}{2}{h^{(1)}}_{ik}{{h^{(1)}}_{j}}^{k})\\
{R}_{ijkl}=-\frac{t^2}{l^2}W_{ijkl}   \\
{R}_{ijk\rho}=\frac{t^2}{2} C_{ijk} \\
\ea
\ee
where $W$ and $C$ are the Weyl and Cotton tensors.

If $d\ne 4$, we can define ${Z^{(4)}}_{ij}$ as 
\be
R_{\r i \r j}=-\frac{t^{2}l^2}{2}\frac{{Z^{(4)}}_{ij}}{d-4}
\ee
and using the equation (\ref{1order}) for ${h^{(2)}}_{ij}$ we establish that 
\bea
{Z^{(4)}}_{ij}&=&\frac{1}{2l^4}\left({h^{(1)}}_{kl}{h^{(1)}}^{kl}g_{ij}
   -4{h^{(1)}}_{ik}{{h^{(1)}}_{j}}^{k}
   +2\nabla_{k}\nabla_{(i}{{h^{(1)}}_{j)}}^{k}
   - \nabla^{2}{h^{(1)}}_{ij}\right.\nonumber\\
&& \left.  - \nabla_{i}\nabla_{j}{h^{(1)}}  
   -(d-4){h^{(1)}}_{ik}{{h^{(1)}}_{j}}^{k}\right)\ = \
   -\frac{1}{2}B_{ij}
\eea
where $B_{ij}$ is the Bach tensor (in any dimension). Then ${Z^{(4)}}_{ij}$ is
precisely what we have called obstruction in $d=4$.


\par

 The four-dimensional  obstruction has then been obtained afresh 
in a natural geometric way 
 from the ambient Riemann tensor. A natural question is whether this
could be generalized to higher dimensions. 
\par
Let us consider a component of the ambient tensor 
\be \label{nablaR}
\nabla_{\a_{1}}\ldots\nabla_{\a_{n}}{\mathcal R}_{\l\m\n\s}
\ee
 with $n$
indices fixed: $\a_{1}=\ldots=\a_{n}=\r$, and take the $\r \to 0$ limit.
This leads to:
\be
\ba{l}
(\nabla_{\r})^{n}R_{\r i \r j}=-\frac{t^{2}}{2}\ l^{2n+2}\ \frac{{Z^{(4+2n)}}_{ij}}{d-(4+2n)} 
\\
(\nabla_{\r})^{n}{R}_{ijkl}=
    -t^2\  l^{2n-2}\   {\mathcal{W}^{(4+2n)}}_{ijkl}   
\\
 (\nabla_{\r})^{n}{R}_{ijk\rho}=
     \frac{t^2}{2}\ l^{2n}\ {\mathcal{C}^{(4+2n)}}_{ijk} 
\\
\ea
\ee
where $\mathcal{W}^{(4+2n)}$ and $\mathcal{C}^{(4+2n)}$ are generalizations of the Weyl ($W=\mathcal{W}^{(4)}$) and
Cotton ($C=\mathcal{C}^{(4)}$) tensors.

\par

Performing the same calculation as before for $n=1,\ldots$,  we obtain the
obstructions for higher dimensions ( the results are presented in the appendix \ref{obs}).
These tensors yield the Dirichlet obstruction in the corresponding dimension.
\par
From the general expression (\ref{nablaR}) 
we obtain the relationship 
\be\label{recursive}
\left. (\nabla_{\r})^{n+1}{\mathcal R}_{\r i\r j} \right| _{\r =0}
=\left. \partial_{\r}(\nabla_{\r})^{n}{\mathcal R}_{\r i\r j} \right| _{\r=0}
-\left. 2\Gamma_{\r (i|}^{k} (\nabla_{\r})^{n} {\mathcal R}_{\r |j) \r k} \right|_{\r=0}
\ee
(where $\Gamma_{\r i}^{k}=\frac{1}{2}h^{kl}\partial_{\r}h_{il}$ is a
Christoffel symbol of the ambient metric) which relates $Z^{(p)}$ to $Z^{(p-2)}$,
 suggesting a  recurrent computation. 
Unfortunately this is not
straightforward because we need to know the general tensor $\nabla_{\r}\ldots\nabla_{\r}{\mathcal R}_{\l\m\n\s}$ 
not only in the $\r
\to 0$ limit, but also for general $\r$. Nevertheless, using the fact that
$Z^{(4)}$ is traceless, it can be shown inductively from (\ref{recursive}) that $Z^{(4+2n)}$ is traceless as well. Moreover, as
\be
{\nabla_{\r}}^{n)}{\mathcal R}_{\r \r}=
{\partial_{\r}}^{n)}{\mathcal R}_{\r \r}=0
\ee
we see that  the property of $Z^{(4+2n)}$ being tracefree is
equivalent (in the $\r \to 0$ limit) to the trace equations for $h^{(n+2)}$
given by the equation (\ref{EOM3}).

\subsection{The  Bianchi identities}

Let us now study the Bianchi identity in the ambient space $A_{n+2}$
\be
\nabla_{[\m}{\mathcal R}_{\n\l]\s\t}=0
\ee
If we contract, for instance, $\m$ and $\t$, and remember that the ambient space is
Ricci-flat, it reduces to
\be
\nabla_{\m}{{\mathcal R}^{\m}}_{\n\l\s}=\nabla_{\l}{\mathcal R}_{\s\n}-\nabla_{\s}{\mathcal R}_{\l\n}=0
\ee
In the particular case $\n=i$, $\l=\r$, $\s=j$, one gets close to the boundary
 (using (\ref{riemann0}))  the identity
\be
B_{ij}=\nabla^{k}C_{kij}+A^{kl}W_{kijl}
\ee
which is the Bach tensor (i.e. $Z^{(4)}$) in terms of A, C, and W.
\par
We can go further and assume that this is a general feature, that is, the obstruction
could be expressed in terms of the generalization of the Weyl and Cotton tensors. 
\par
Starting from the identity
\be
\nabla_{\r}\nabla_{\m}{{\mathcal R}^{\m}}_{i\r j}=0
\ee
we can perform a calculation along similar lines. The result however 
is more complicated than expected:
\be\ba{l}
{Z^{(6)}}_{ij}=\frac{1}{2}\left[ \nabla^{k}{\mathcal{C}^{(6)}}_{kij} +
  A^{kl}{\mathcal{W}^{(6)}}_{kijl}-\frac{1}{2}A^{kl}\nabla_{k}{C}_{lij}-\frac{1}{2}A^{kl}{A_{k}}^{s}W_{lijs}-\frac{1}{d-4}{A_{i}}^{s}B_{js}\right. \\
\qquad \left. +\frac{1}{d-4}A B_{ij}+\frac{1}{2}C_{ilk}{C^{lk}}_{j}+\frac{1}{2}C_{ikl}{C_{j}}^{lk}-\frac{1}{2}\frac{1}{d-4}B^{kl}W_{kijl}\right]\\
\ea\ee


\section{The $d=8$ Holographic Anomaly}
As a by-product of this computation, the holographic anomaly in $d=8$ can be evaluated. This anomaly, 
when integrated, would give rise to $Z^{(8)}$ upon metric variation.
 It should correspond to the anomaly of a conformal invariant quantum field theory 
 and, as such, 
expressible in terms of the Euler characteristic $E_8$ and a basis of conformal invariants, $W_8$. We are not
aware, however, of any existing explicit construction of such a basis, so that we express our result
in terms of Riemann tensors, their derivatives, and contracions thereof. This yields:

\bea
a^{(8)}&=&\frac{1}{32}\left\{               
   \frac{1}{54}R^{i j}R_{i j}R^{k l}R_{k l} 
   -\frac{11}{756}R^{2}R^{i j}R_{i j}
   +\frac{17}{18522}R^{4}
   +\frac{1}{27}R_{k i j l}R^{k l}R^{i p}{R^{j}}_{p}\right.\nonumber \\
&&   -\frac{1}{18}RR_{k i j l}R^{k l}R^{i j}
   -\frac{1}{12}R_{k p q l}R^{i p q j}R^{k l}R_{i j}
   -\frac{13}{5292}R^{2}\nabla^{2}R
   -\frac{43}{5292}RR^{i j}\nabla_{i}\nabla_{j}R\nonumber \\
&& +\frac{7}{216}RR^{i j}\nabla^{2}R_{i j} 
   +\frac{1}{126}R^{i j}R_{i j}\nabla^{2}R
   -\frac{1}{27}R^{i j}R^{k l}\nabla_{i}\nabla_{j}R_{k l}
   +\frac{1}{27}R^{i j}R^{k l}\nabla_{k}\nabla_{i}R_{j l}\nonumber \\
&& +\frac{1}{189}R^{i l}{R^{j}}_{l}\nabla_{i}\nabla_{j}R     
   +\frac{1}{36}R^{i j}R^{k l}\nabla^{2}R_{k i j l}
   -\frac{1}{42}R^{i j}R_{k i j l}\nabla^{k}\nabla^{l}R
    +\frac{1}{12}R^{i j}R_{k i j l}\nabla^{2}R^{k l}\nonumber  \\ 
&& -\frac{1}{5292}R\nabla_{k}R\nabla^{k}R
   +\frac{1}{216}R\nabla_{k}R_{i j}\nabla^{k}R^{i j}
   +\frac{1}{1008}R\nabla^{2}\nabla^{2}R
   +\frac{1}{168}R^{i j}\nabla^{2}\nabla_{i}\nabla_{j}R\nonumber  \\
&& -\frac{1}{72}R^{i j}\nabla^{2}\nabla^{2}R_{i j}
   -\frac{1}{882}R^{i j}\nabla_{i}R\nabla_{j}R
   +\frac{1}{189}R^{i j}\nabla_{i}R_{j k}\nabla^{k}R
   +\frac{1}{54}R^{i j}\nabla_{i}R_{k l}\nabla_{j}R^{k l}\nonumber  \\
&& +\frac{1}{27}R^{i j}\nabla_{k}R_{i l}\nabla^{l}{R^{k}}_{j}
   -\frac{1}{27}R^{i j}\nabla_{i}R_{k l}\nabla^{k}{R^{l}}_{j}
   +\frac{1}{18}R^{i j}\nabla_{p}R_{k i j l}\nabla^{p}R^{k l}
   +\frac{1}{3528}\nabla^{2}R\nabla^{2}R \nonumber  \\
&& \left. -\frac{1}{784}\nabla_{i}\nabla_{j}R\nabla^{i}\nabla^{j}R
   +\frac{1}{168}\nabla_{i}\nabla_{j}R\nabla^{2}R^{i j}
   -\frac{1}{144}\nabla^{2}R^{i j}\nabla^{2}R_{i j} \right\}
\eea

After this work was completed, we found a previous proposal for $a^{(8)}$ in the context of the AdS/CFT correspondence (\cite{odintsov}), although from a slightly different approach.
It is worth noting that, by construction, this expression depends only on
the coefficients $h^{n}$ which in turn are given in terms of $h^{1}$ (due to
the recurrence relation stemming from the equations of motion). Then, as has been
already remarked in (\ref{h1}), the holographic 
anomaly vanishes in a Ricci-flat background, and this is a
general property of the conformal anomaly at any dimension obtained by the holographic setup.

\par
\section{Conclusions}
The higher dimensional ($6\leq d\leq 10$)  Dirichlet obstruction $Z^{(d)}$ have been computed 
in several 
 ways,
which we believe are equivalent. 
\par
The use of the so called {\em ambient space}, with a codimension two
associated space-time
has proved very fruitful both in computing the obstructions themselves as well as in 
uncovering unexpected relationships between the obstructions in different dimensions.
\par
We have also analyzed the relationship of the generic obstruction to the generic conformal anomaly (
for conformally invariant matter), along the lines of \cite{haro},
although here clearly we have only scratched the surface of a deep problem, which deserves further study.
\par
The full expression of the eight dimensional holographic anomaly  has been
derived in terms of geometrical quantities .
It would be most interesting to check (\cite{vandeVen}) 
the expressions of the holographic anomaly in dimensions six and eight.
This would imply, in particular, to verify that the total coefficient of all independent Riemann contractions 
vanishes, due to the basic property of the holographic anomaly that it is zero for Einstein spaces.

\section*{Acknowledgments}
One of us (E.A.) acknowledges useful correspondence with Stanley Deser. 
This work ~~has been partially supported by the
European Commission (HPRN-CT-200-00148) and CICYT (Spain).      


\begin{appendix}

\section{Bach-flat spaces}
An interesting question is the characterization of those spaces for which 
there is no obstruction. In $d=4$ dimensions, this is equivalent to find Bach-flat spaces.
It is known (cf.\cite{kozameh}) that all conformally
Einstein spaces are in this class, but there are other solutions besides
those (cf. \cite{Schmidt}), and a complete characterization is not known to us.
\par
The non-asymptotically flat solutions found by Schmidt are a generalization of
well-known Kasner solutions, with metric
\be
ds^2 = dt^2  - \sum_{i=1}^3a^2_i  dx_i^2
\ee
with Hubble parameters $h_i = a_i^{-1} da_i/dt$, $h=\Sigma h_i$
and anisotropy parameters $m_i = h_i-   h/3$
If we denote
$r = (h^2_1 + h_1h_2 + h_2^2)^{1/2}$  and $p = h_1/r$, 
 them the vanishing of Bach's tensor is equivalent to:                
\be
3d^2(pr)/dt^2 =  8pr^3 + 4c, \quad c = {\rm const.,} \quad   p^2 \le  4/3
\,  ,  
\ee
\be 
9(dp/dt)^2 r^4 =   [2r d^2 r/dt^2 - (dr/dt)^2 - 4r^4] (4r^2 - 3p^2r^2) \,
. 
\ee
 Each solution of this
 system of equations with $dp/dt \ne  0$  represents a non asymptotically flat
metric enjoying vanishing Bach tensor.


\section{Higher order equations of motion}\label{HighEOMS}

In this appendix can be found the equations of motion (\ref{EOM1}) and
(\ref{EOM3}), from order $\rho^2$ to order $\rho^4$. ${R^{(n)}}_{ij}$ corresponds to the power expansion of the Ricci tensor
$R_{ij}[h]$, and we employ the notation ${D{h^{(n)}}_{a b}}^{c}=\nabla_{a}{{h^{(n)}}_{b}}^{c}+\nabla_{b}{{h^{(n)}}_{a}}^{c}-\nabla^{c}{{h^{(n)}}_{a b}}$.
 
Order $\r^2$ 
\bea \label{2order}
&&\bullet\quad 3( d-6){h^{(3)}}_{ij}
      + \left( 3h^{(3)} -3{h^{(1)}}_{kl}{h^{(2)}}^{kl}+
      {{h^{(1)}}^{k}}_{s}{{h^{(1)}}^{s}}_{l}{{h^{(1)}}^{l}}_{k}\right) g_{ij}
      + 8{{h^{(2)}}^{m}}_{(i}{h^{(1)}}_{j)m} \nonumber \\
&&\quad -2{h^{(1)}}_{il}{h^{(1)}}^{lm}{h^{(1)}}_{mj}- h^{(1)}{h^{(2)}}_{ij} 
      -l^2 {R^{(2)}}_{ij}=0     \\
&&\bullet\quad  12h^{(4)} -9{h^{(1)}}^{kl}{h^{(3)}}_{kl}-4{h^{(2)}}^{kl}{h^{(2)}}_{kl} +7
      {h^{(1)}}^{ik}{{h^{(1)}}^{j}}_{k}{h^{(2)}}_{ij} \nonumber\\
&&  \quad -\frac{3}{2}{{h^{(1)}}^{i}}_{j}{{h^{(1)}}^{j}}_{k}{{h^{(1)}}^{k}}_{l}{{h^{(1)}}^{l}}_{i} =0 
\eea
Order $\r^3$ 
\bea \label{3order}
&&\bullet\quad
4(d-8){h^{(4)}}_{ij} +\left( 4h^{(4)}-4
      {h^{(3)}}^{kl}{h^{(1)}}_{kl}-2{h^{(2)}}_{kl}{h^{(2)}}^{kl}
     +4{h^{(2)}}_{kl}{h^{(1)}}^{kn}{{h^{(1)}}^{l}}_{n} \right.\nonumber\\ 
&&  \quad  \left. -{h^{(1)}}_{kl}{h^{(1)}}^{ln}{{h^{(1)}}_{n}}^{m}
     {{h^{(1)}}_{m}}^{l}  \right)g_{ij}  
     -2{h^{(1)}}{h^{(3)}}_{ij}+12{h^{(3)}}_{k(i}
     {{h^{(1)}}_{j)}}^{k}+\left(-2{h^{(2)}}\right. \nonumber\\
&&  \quad \left. +{h^{(1)}}^{kl}{h^{(1)}}_{kl}\right)   
     {h^{(2)}}_{ij} +8{h^{(2)}}_{ik}{{h^{(2)}}_{j}}^{k}-8{h^{(2)}}_{l(i}
      {h^{(1)}}_{j)k}{h^{(1)}}^{kl} 
     -2{h^{(2)}}^{kl}{h^{(1)}}_{ik}{h^{(1)}}_{jl} \nonumber\\
&&  \quad  +2{h^{(1)}}_{ik}
    {h^{(1)}}_{jl}{h^{(1)}}^{kn}{{h^{(1)}}_{n}}^{l}
      - l^{2}{R^{(3)}}_{ij}= 0 \\
&&\bullet\quad  20{h^{(5)}}-16{h^{(4)}} ^{kl}{h^{(1)}}_{kl}-14 {h^{(3)}}^{kl}{h^{(2)}}_{kl} 
    +13{{h^{(3)}}^{i}}_{j}{{h^{(1)}}^{j}}_{k}{{h^{(1)}}^{k}}_{i}   
    +12{{h^{(2)}}{i}}_{j}{{h^{(2)}}^{j}}_{k}{{h^{(1)}}^{k}}_{i} \nonumber\\
&& \quad -11 {{h^{(1)}}^{i}}_{j}{{h^{(1)}}^{j}}_{k}{{h^{(1)}}^{k}}_{l}{{h^{(2)}}^{l}}_{i}
    +2{{h^{(1)}}^{i}}_{j}{{h^{(1)}}^{j}}_{k}{{h^{(1)}}^{k}}_{l}{{h^{(1)}}^{l}}_{m}{{h^{(1)}}^{m}}_{i}
\eea 
Order $\r^4$ 
\bea \label{4order}
&&\bullet\quad
5(d-10){h^{(5)}}_{ij}+\left( 5h^{(5)}- 5 {h^{(4)}}^{kl} {h^{(1)}}_{kl} 
    -  5 {h^{(3)}}^{kl} {h^{(2)}}_{kl} +
      5 {h^{(3)}}^{kl}{ {h^{(1)}}_{k}}^{n}{ {h^{(1)}}_{ln}}\right.\nonumber\\ 
&&  \quad  + 5 {h^{(2)}}^{kl}{ {h^{(2)}}_{k}}^{n}{ {h^{(1)}}_{ln}}
    \left.-5 {h^{(1)}}^{kl}{{h^{(1)}}_{k}}^{n}{{h^{(1)}}_{n}}^{m}
    {{h^{(2)}}_{ml}}+ {h^{(1)}}^{kl}{{h^{(1)}}_{k}}^{n}
     {{h^{(1)}}_{n}}^{m}{{h^{(1)}}_{m}}^{p}{{h^{(1)}}_{pl}} \right) 
      g_{ij}\nonumber\\ 
&&  \quad   -3{h^{(1)}}{h^{(4)}}_{ij}+16{h^{(1)}}_{k(i}{{h^{(4)}}_{j)}}^{k}
   + \left(-4 {h^{(2)}}+2 {h^{(1)}}^{kl}{h^{(1)}}_{kl}\right)
    {h^{(3)}}_{ij} +24 {h^{(3)}}_{k(i}{{h^{(2)}}_{j)}}^{k}\nonumber\\  
&& \quad  -12 {h^{(3)}}_{k(i}{h^{(1)}}_{j)l}  {h^{(1)}}^{kl}
   -2 {h^{(3)}}^{kl}{h^{(1)}}_{ik}{h^{(1)}}_{jl} 
   + \left( -3 {h^{(3)}} +3 {h^{(2)}}^{kl} {h^{(1)}}_{kl}\right.  \nonumber\\
&& \quad \left.-{h^{(1)}}^{kn}{{h^{(1)}}^{l}}_{n}{h^{(1)}}_{kl}\right) {h^{(2)}}_{ij}
     -8 {h^{(2)}}_{ik} {h^{(2)}}_{lj} {h^{(1)}}^{kl}  
     - 8 {h^{(2)}}^{kl}{h^{(2)}}_{k(i} {h^{(1)}}_{j)l} \nonumber \\
 &&  \quad +8{h^{(1)}}^{kn}{{h^{(1)}}^{l}}_{n}{h^{(1)}}_{k(i}
   {h^{(2)}}_{j)l} 
    +4{h^{(2)}}^{kl}{{h^{(1)}}^{l}}_{n}{h^{(1)}}_{ik}{h^{(1)}}_{jl}   
   -2 {h^{(1)}}^{kn}{h^{(1)}}^{lm}{h^{(1)}}_{mn}{h^{(1)}}_{ik}
    {h^{(1)}}_{jl}\ \nonumber\\
&& \quad  -l^2 {R^{(4)}}_{ij}=0   \\
&&\bullet\quad 30 h^{(6)}+ \ldots = 0 
\eea
The corresponding coefficients of the $R[h]$ expansion are
\begin{eqnarray}
{R^{(2)}}_{i j}&=& \frac{1}{2}
   \left[ \nabla_{k} \left( {D{h^{(2)}}_{i j}}^{k} - 
       {{h^{(1)}}^{k}}_{l}{D{h^{(1)}}_{i j}}^{l}  \right) -
  \nabla_{i}\left(\nabla_{\n}{h^{(2)}} - 
    {{h^{(1)}}^{k}}_{l}{D{h^{(1)}}_{k j}}^{l} \right)\right.\nonumber \\
 & & \left.  +  \frac{1}{2}\left( 
    {D{h^{(1)}}_{i j}}^{k} {D{h^{(1)}}_{k l}}^{l}  
   - {D{h^{(1)}}_{i k}}^{l}{D{h^{(1)}}_{j l}}^{k} \right)
  \right] \\
{R^{(3)}}_{i j}&=&\frac{1}{2}\left[\nabla_{k}\left(
    {D{h^{(3)}}_{ij}}^{k} - {{h^{(1)}}^{k}}_{l}{D{h^{(2)}}_{i j}}^{l}
    - [ {{h^{(2)}}^{k}}_{l}- {h^{(1)}}^{km}{h^{(1)}}_{lm}]
    {D{h^{(1)}}_{i j}}^{l}\right)\right.\nonumber\\ 
 & &   -\nabla_{i}\left(\nabla_{j} {h^{(3)}}  
     - {{h^{(1)}}^{k}}_{l}{D{h^{(2)}}_{kj}}^{l} -
   [{{h^{(2)}}^{k}}_{l}- {h^{(1)}}^{km}{h^{(1)}}_{lm}  ]
    {D{h^{(1)}}_{k j}}^{l}  \right)\nonumber\\  
  & & +\frac{1}{2}\left({D{h^{(1)}}_{i j}}^{k}
    {D{h^{(2)}}_{k l}}^{l}+{D{h^{(2)}}_{ij}}^{k}{D{h^{(1)}}_{k l}}^{l}
   -2 {D{h^{(1)}}_{k (i}}^{l}{D{h^{(2)}}_{j) l}}^{k}\right) \nonumber\\
  & & \left. +\frac{1}{2}{{h^{(1)}}^{k}}_{l}\left(
    2 {D{h^{(1)}}_{k (i}}^{m}{D{h^{(1)}}_{j) m}}^{l}  
    - {D{h^{(1)}}_{ij}}^{l}{D{h^{(1)}}_{km}}^{m} 
    - {D{h^{(1)}}_{i j}}^{m}
   {D{h^{(1)}}_{m k}}^{l}\right) \right]\\
{R^{(4)}}_{i j}&=&\frac{1}{2}\left[\nabla_{k}
  \left( {D{h^{(4)}}_{ij}}^{k} -  {{h^{(3)}}^{k}}_{l}
    {D{h^{(1)}}_{i j}}^{l} - {{h^{(1)}}^{k}}_{l}{D{h^{(3)}}_{i j}}^{l}
    - [ {{h^{(2)}}^{k}}_{l}- {h^{(1)}}^{km}{h^{(1)}}_{lm}  ]
    {D{h^{(2)}}_{i j}}^{l} \right.\right.\nonumber\\
&  &\left. + [{{h^{(1)}}^{km}}{h^{(2)}}_{lm}  +  {{h^{(2)}}^{km}}{h^{(1)}}_{lm}
    - {h^{(1)}}^{km}{{h^{(1)}}_{m}}^{n}{h^{(1)}}_{ln}]
    {D{h^{(1)}}_{i j}}^{l}  \right)\nonumber\\
& &  -\nabla_{i}\left( \nabla_{j}{h^{(4)}}-{{h^{(3)}}^{k}}_{l}
   {D{h^{(1)}}_{k j}}^{l} -{{h^{(1)}}^{k}}_{l}{D{h^{(3)}}_{k j}}^{l}
   -  [{{h^{(2)}}^{k}}_{l}- {h^{(1)}}^{km}{h^{(1)}}_{lm}  ]  
   {D{h^{(2)}}_{k j}}^{l} \right.\nonumber\\
& & \left. + [{{h^{(1)}}^{km}}{h^{(2)}}_{lm}  +  
   {{h^{(2)}}^{km}}{h^{(1)}}_{lm}
    - {h^{(1)}}^{km}{{h^{(1)}}_{m}}^{n}{h^{(1)}}_{ln}]
    {D{h^{(1)}}_{k j}}^{l} \right)\nonumber\\
& &  \frac{1}{2}\left( {D{h^{(3)}}_{i j}}^{k}
    {D{h^{(1)}}_{k l}}^{l}+{D{h^{(1)}}_{ij}}^{k}{D{h^{(3)}}_{k l}}^{l}
   -2 {D{h^{(3)}}_{k (i}}^{l}{D{h^{(1)}}_{j) l}}^{k} 
   + {D{h^{(2)}}_{i j}}^{k}{D{h^{(2)}}_{k l}}^{l} \right.\nonumber\\ 
& & \left.   + {D{h^{(2)}}_{i k}}^{l}{D{h^{(2)}}_{j l}}^{k} \right) 
   +\frac{1}{2}\left({{h^{(2)}}^{k}}_{l} -
  {h^{(1)}}^{k n}{h^{(1)}}_{l n}\right)\left(
    2 {D{h^{(1)}}_{k (i}}^{m}{D{h^{(1)}}_{j) m}}^{l}\right.  \nonumber\\
& & \left.
    - {D{h^{(1)}}_{ij}}^{l}{D{h^{(1)}}_{km}}^{m} 
    - {D{h^{(1)}}_{i j}}^{m}{D{h^{(1)}}_{m k}}^{l}\right)\nonumber\\
& &   +\frac{1}{2}\left(
    {{h^{(1)}}^{k}}_{m}{{h^{(1)}}^{l}}_{n}\right)
    \left( {D{h^{(1)}}_{ij}}^{m}{D{h^{(1)}}_{kl}}^{n} 
  - {D{h^{(1)}}_{ik}}^{n}{D{h^{(1)}}_{jl}}^{m}   \right)\nonumber\\
& & +\frac{1}{2}{{h^{(1)}}^{k}}_{l}\left(
    2 {D{h^{(2)}}_{k (i}}^{m}{D{h^{(1)}}_{j) m}}^{l}  
    - {D{h^{(2)}}_{ij}}^{l}{D{h^{(1)}}_{km}}^{m} 
    - {D{h^{(2)}}_{i j}}^{m}{D{h^{(1)}}_{m k}}^{l}\right)\nonumber\\
& &  \left. +\frac{1}{2}{{h^{(1)}}^{k}}_{l}\left(
    2 {D{h^{(1)}}_{k (i}}^{m}{D{h^{(2)}}_{j) m}}^{l}  
    - {D{h^{(1)}}_{ij}}^{l}{D{h^{(2)}}_{km}}^{m} 
    - {D{h^{(1)}}_{i j}}^{m}{D{h^{(2)}}_{m k}}^{l}\right)\right]
\end{eqnarray}

\section{The Dirichlet obstructions in higher dimensions}\label{obs}
Following the procedure explained in section \ref{C2}, we present in this
appendix the obstructions in higher dimension. We employ the Ricci expansion
coefficients ${R^{(n)}}_{ij}$ defined in the previous Appendix. 

For $n=1$ the six-dimensional ($d=6$) obstruction is given by the tensor
\bea
l^6{Z^{(6)}}_{ij}&=&\left(2{h^{(1)}}_{kl}{h^{(2)}}^{kl}
   -{h^{(1)}}^{kl}{{h^{(1)}}_{l}}^{m}{{h^{(1)}}_{mk}}\right)g_{ij}
   -16{h^{(2)}}_{k(i}{{h^{(1)}}_{j)}}^{k}
   +4{h^{(1)}}^{kl}{h^{(1)}}_{il}{h^{(1)}}_{jk} \nonumber\\
&& +2{h^{(1)}}{h^{(2)}}_{ij}
   +2l^{2}{R^{(2)}}_{ij} 
   +(d-6)\left(-4{h^{(2)}}_{k(i}{{h^{(1)}}_{j)}}^{k}
   +{h^{(1)}}^{kl}{h^{(1)}}_{il}{h^{(1)}}_{jk}\right)
\eea

For $n=2$ we get 
\bea
l^8{Z^{(8)}}_{ij}&=&\left(6{h^{(1)}}_{kl}{h^{(3)}}^{kl}
   +4{h^{(2)}}_{kl}{h^{(2)}}^{kl}
   -10{h^{(2)}}^{kl}{{h^{(1)}}_{l}}^{m}{h^{(1)}}_{mk}\right)g_{ij}
   + 12 {h^{(1)}}{h^{(3)}}_{ij}\nonumber\\
&&  +\left(12{h^{(2)}}
   -6{h^{(1)}}_{kl}{h^{(1)}}^{kl}\right){h^{(2)}}_{ij}
   -72{h^{(3)}}_{k(i}{{h^{(1)}}_{j)}}^{k}
   -48{h^{(2)}}_{il}{{h^{(2)}}_{j}}^{l} \nonumber\\
&& +48{h^{(2)}}_{k(i}{{h^{(1)}}_{j)l}}{h^{(1)}}^{kl}
   +12{h^{(2)}}^{kl}{h^{(1)}}_{jk}{h^{(1)}}_{il}   
   -12{h^{(1)}}^{km}{{h^{(1)}}_{m}}^{l}{h^{(1)}}_{ik}{h^{(1)}}_{jl}\nonumber\\
&&  +6l^{2}{R^{(3)}}_{ij} 
    +(d-8)\left(-18{h^{(3)}}_{k(i}{{h^{(1)}}_{j)}}^{k}
   -8{h^{(2)}}_{il}{{h^{(2)}}_{j}}^{l}
   +4{h^{(2)}}^{kl}{h^{(1)}}_{jk}{h^{(1)}}_{il}\right.\nonumber\\
&&   +10{h^{(2)}}_{k(i}{{h^{(1)}}_{j)l}}{h^{(1)}}^{kl}
 \left.-3{h^{(1)}}^{km}{{h^{(1)}}_{m}}^{l}{h^{(1)}}_{ik}{h^{(1)}}_{jl}\right)
\eea
\par

Finally, for $n=3$ the result is
\bea
l^{10}{Z^{(10)}}_{ij}&=&\left( 24 {h^{(1)}}_{kl}{h^{(4)}}^{kl} 
   +36{h^{(1)}}_{kl}{h^{(3)}}^{kl}
   - 42 {h^{(1)}}^{kl}{{h^{(1)}}_{l}}^{m}{h^{(3)}}_{mk} 
   -48 {h^{(2)}}^{kl}{{h^{(2)}}_{l}}^{m}{h^{(1)}}_{mk}\right.\nonumber\\ 
&& \left.+54 {h^{(1)}}^{kl}{{h^{(1)}}_{l}}^{m}{{h^{(1)}}_{m}}^{n}{h^{(2)}}_{nk}
   -12 {h^{(1)}}^{kl}{{h^{(1)}}_{l}}^{m}{{h^{(1)}}_{m}}^{n}{{h^{(1)}}_{n}}^{p}
   {h^{(1)}}_{pk} \right) g_{ij} \nonumber\\
&&   +72{h^{(1)}}{h^{(4)}}^{ij}
   -384 {h^{(1)}}_{k(i}{h^{(4)}_{j)}}^{k} - 24{h^{(3)}}_{ij} 
   \left(-4{h^{(2)}} 
   + 2{h^{(1)}}_{kl}{h^{(1)}}^{kl} \right) \nonumber\\
&&   -576 {h^{(3)}}_{k(i}{h^{(2)}_{j)}}^{k}\
   +288{h^{(3)}}_{k(i}{h^{(1)}}_{j)l}{h^{(1)}}^{kl}  
   +48 {h^{(3)}}^{kl}{h^{(1)}}_{ik}{h^{(1)}}_{jl}\nonumber\\
&&    -24 {h^{(2)}}_{ij}\left( 3{h^{(3)}} -3{h^{(1)}}_{kl}{h^{(2)}}^{kl}
    +{h^{(1)}}^{kl}{{h^{(1)}}_{l}}^{m}{h^{(1)}}_{mk}\right) 
   +192{h^{(2)}}_{ik}{{h^{(2)}}_{jl}}{h^{(1)}}^{kl}\nonumber\\
&&   +192{h^{(2)}}^{kl}{{h^{(2)}}_{k(i}}{h^{(1)}}_{j)l} 
    -192{h^{(1)}}^{kn}{{h^{(1)}}^{l}}_{n}{h^{(2)}}_{k(i}{h^{(1)}}_{j)l}
   \nonumber\\ 
&&  -96{h^{(2)}}^{kn}{{h^{(1)}}^{l}}_{n}{h^{(1)}}_{k(i}{h^{(1)}}_{j)l}
  +48{h^{(1)}}^{kn}{{h^{(1)}}^{lm}}{{h^{(1)}}_{nm}}{h^{(1)}}_{ki}{h^{(1)}}_{jl}   + 24 l^{2}{R^{(4)}}_{ij} \nonumber\\
&&   + (d-10) \left(-96{h^{(1)}}_{k(i}{{h^{(4)}}_{j)}}^{k} 
   -84{h^{(3)}}_{k(i} {{h^{(2)}}_{j)}}^{k}\right. 
   +57{h^{(3)}}_{k(i}{h^{(1)}}_{j)l}{h^{(1)}}^{kl}\nonumber\\ 
&&   +21 {h^{(3)}}^{kl}{h^{(1)}}_{ik}{h^{(1)}}_{jl} 
   +44{h^{(2)}}^{kl}{{h^{(2)}}_{k(i}}{h^{(1)}}_{j)l}
   + 28 {h^{(2)}}_{ik}{{h^{(2)}}_{jl}}{h^{(1)}}^{kl}\nonumber\\
&& -37{h^{(1)}}^{kn}{{h^{(1)}}^{l}}_{n}{h^{(2)}}_{k(i}{h^{(1)}}_{j)l}
     -29 {h^{(1)}}_{k(i}{{h^{(4)}}_{j)}}^{k} 
   -84{h^{(3)}}_{k(i} {{h^{(2)}}_{j)}}^{k} \nonumber\\  
&&    \left.+12{h^{(1)}}^{kn}{{h^{(1)}}^{lm}}{{h^{(1)}}_{nm}}{h^{(1)}}_{ki}
   {h^{(1)}}_{jl} \right)
\eea

\end{appendix}


\end{document}